\documentclass{IEEEtran}
\IEEEoverridecommandlockouts
\usepackage{cite}
\usepackage{amsmath,amssymb,amsfonts}
\usepackage{graphicx}
\usepackage{siunitx}
\usepackage{subcaption}
\usepackage{booktabs}

\usepackage{amsthm}
	\theoremstyle{remark}
		\newtheorem{rmk}{Remark}

\begin{document}

\title{A Hybrid Modelling of a Water and Air Injector in a Subsonic Icing Wind Tunnel
	\thanks{This work was supported by the Dispositif Recherche of M\'etropole Rouen Normandie through the COPOGIRT project.
	For this reason and the purpose of Open Access, the authors have applied a CC BY public copyright licence to any Author Accepted Manuscript (AAM) version arising from this submission.}
}

\author{C\'esar Hern\'andez-Hern\'andez, Thomas Chevet, Rihab el Houda Thabet, and Nicolas Langlois
\thanks{The authors are with Universit\'e de Rouen Normandie, ESIGELEC, IRSEEM, 76000 Rouen, France
	{\tt\small \{cesar.hernandez, thomas.chevet, rihab.hajrielhouda, nicolas.langlois\}@esigelec.fr}}%
}

\maketitle

\begin{abstract}
	The study of droplet generation in wind tunnels in conducting icing experiments is of great importance in determining ice formation on structures or surfaces, where parameters such as Liquid Water Content (LWC) and Median Volumetric Diameter (MVD) play a relevant role. 
	The measurement of these parameters requires specialised instrumentation. 
	In this paper, several experiments have been carried out in a subsonic wind tunnel facility to study the parameters that are part of the icing process in structures. 
	Furthermore, a mathematical modelling of the constituent subsystems of the plant study that allow us to have a comprehensive understanding of the behaviour of the system is developed using techniques based on first principles and machine learning techniques such as regression trees and neural networks. 
	The simulation results show that the implementation of the model manages to obtain prominent expected values of LWC and MVD within the range of values obtained in the real experimental data. 
\end{abstract}


	\section{Introduction}
		\label{section:introduction}

In the process of structural icing, many parameters are involved, for instance, temperature, wind velocity, air pressure, humidity, liquid water content (LWC) and median volumetric diameter of water droplets (MVD). 
LWC is normally expressed as the number of grammes of liquid water per cubic meter of air. 
It represents the amount of supercooled water droplets that can impact the aircraft surface in a given air mass. 
The diameter of the water droplets is usually characterised as the median volumetric diameter normally expressed in micrometre, and LWC and MVD are closely related \cite{li2022aircraft}. 
Determining the LWC detection of the distribution of droplets produced by a spray system in a wind tunnel is of great importance to prevent ice formation. 
MVD is usually measured with different instruments, which generally require interaction with the droplets under sensitive conditions and are prone to fail under very icy conditions.
Therefore, LWC and MVD are important factors in structural icing. 

The possibility of ice formation under atmospheric conditions has become a major problem in different areas, such as in electrical networks \cite{davis2014forecast}, wind turbines \cite{dierer2011wind,martini2021review,molinder2019use}, aircrafts \cite{gent2000aircraft,cao2015aircraft} and helicopters \cite{kreeger2015progress}. 

Considerable work has been done to study the physics and nature of ice formation \cite{dobesch2005physical}, where LWC and MVD are highly essential parameters related to the occurrence of ice \cite{yang2015coupled}. 
In \cite{knop2021comparison}, the authors integrate three measurement devices into a wind tunnel to measure particle size distribution (PSD), MVD, and LWC. 
Moreover, in \cite{rydblom2020measurement} a $k$-nearest neighbour-based model of ice intensity has been developed from wind velocity, LWC and MVD measurements from two instruments.
In \cite{davison2017droplet}, variables related to LWC are studied to determine the effect on engine conditions due to the change in LWC and atmospheric conditions. 

In this work, we propose a hybrid nonlinear mathematical model of a subsonic icing wind tunnel and its constituent subsystems. 
The model is developed and implemented in Matlab/Simulink \cite{MATLAB:2022}, and obtains values that are within the range of experimental test values, where there are atmospheric conditions measurements that allow LWC values between $0.3$ and $\SI{2.5}{\gram\per\meter\cubed}$ and MVD between $16.9$ and $\SI{48.8}{\micro\meter}$. 
To develop the hybrid model, we use models based on the lumped parameter approach \cite{krosel1986lumped} and data-driven models for the variables involved in the process. 
We show simulation results where different scenarios are considered to study the variables of interest LWC and MVD.

The remainder of the paper is organised as follows. Section \ref{section:setup} presents the information about the experimental plant. 
Section \ref{section:modelling_methodology} presents the mathematical modelling of the plant. 
Section \ref{section:results} presents results obtained with the plant modelling implemented in simulation.
Finally, Section \ref{section:conclusions} presents conclusions and future work.

	\section{Experimental setup}
		\label{section:setup}

In this work, we study an experimental plant consisting in a closed-loop subsonic wind tunnel associated with a water and air injection system.
Before describing the modelling process, we introduce, in this section, information about this plant.
To do so, we give a brief presentation of the two constituent parts.

The first part of the facilities is the water and air injection system.
This system is composed of a water and an air tank, each connected to 12 conduits.
Each of the 24 conduits possesses its control valve to regulate the flow.
Finally, each water conduit is paired with an air conduit into a nozzle that injects a mixture of water and air into the second part of the facilities.

The second part is the wind tunnel itself.
This tunnel is associated with a cooling chamber in order to maintain a negative temperature during experiments.
These negative temperatures coupled with the injection system described before aim to generate an ice fog inside the wind tunnel to test defrost equipment.
Control panels, allowing to adjust various parameters in the plant, are connected to it.

\begin{figure}[!tb]
  \centering
  \includegraphics[width=0.9\columnwidth]{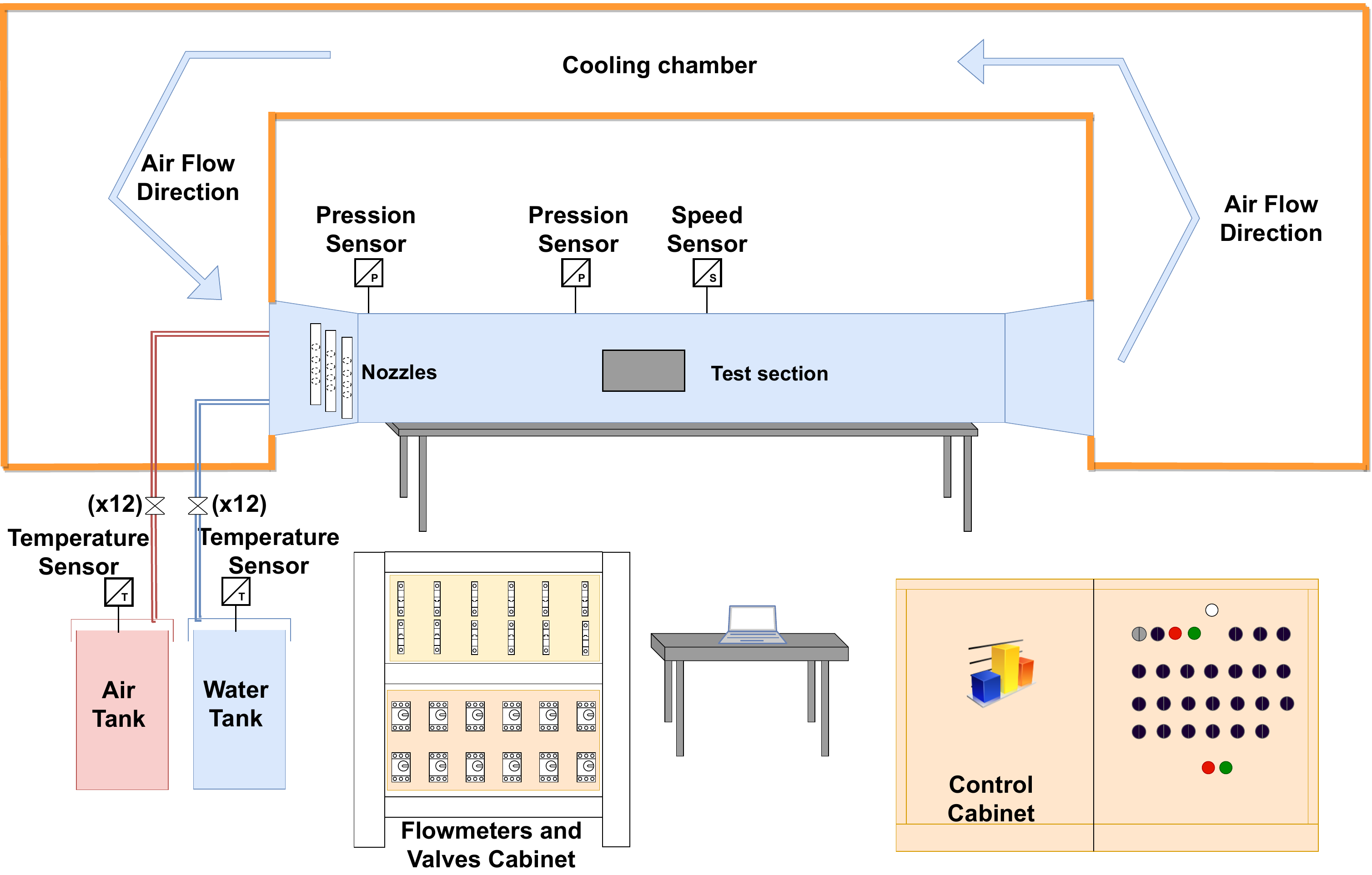}
  \caption{Testbed scheme.}
  \label{fig:testbed}
\end{figure}

A schematic view of this plant is given in Figure \ref{fig:testbed}.

		\subsection{Equipment description}
			\label{subsection:equipment_description}

After this brief introduction, we now provide more details on the equipment of the icing wind tunnel facility.

The \emph{water tank} is considered to be a closed cylindrical tank $\SI{1}{\meter}$ high with a radius of approximately $\SI{20}{\centi\meter}$.
It has a capacity of $\SI{125}{\liter}$.
This tank is pressurised by constant air injection at a pressure of $\SI{7}{\bar}$, and is heated by five resistors (four of $\SI{500}{\watt}$ and one of $\SI{300}{\watt}$).
In addition, a pressure regulator valve is installed for safety purposes.
In terms of sensors, the tank is equipped with a type J thermocouple, which is capable of measuring temperatures between $\num{-40}$ and $\SI{375}{\degreeCelsius}$.
Finally, a sensor is present to measure the water level inside the tank.
For its part, \emph{air tank} is assumed to work as a blower, that is, the air coming from a pressurised air line is heated at a constant chosen temperature and injected into the system at a pressure of $\SI{7}{\bar}$.

\begin{figure}[!tb]
  \begin{minipage}{0.48\columnwidth}
	\centering
	\begin{subfigure}[b]{1\textwidth}
		\centering
		\includegraphics[width=\textwidth]{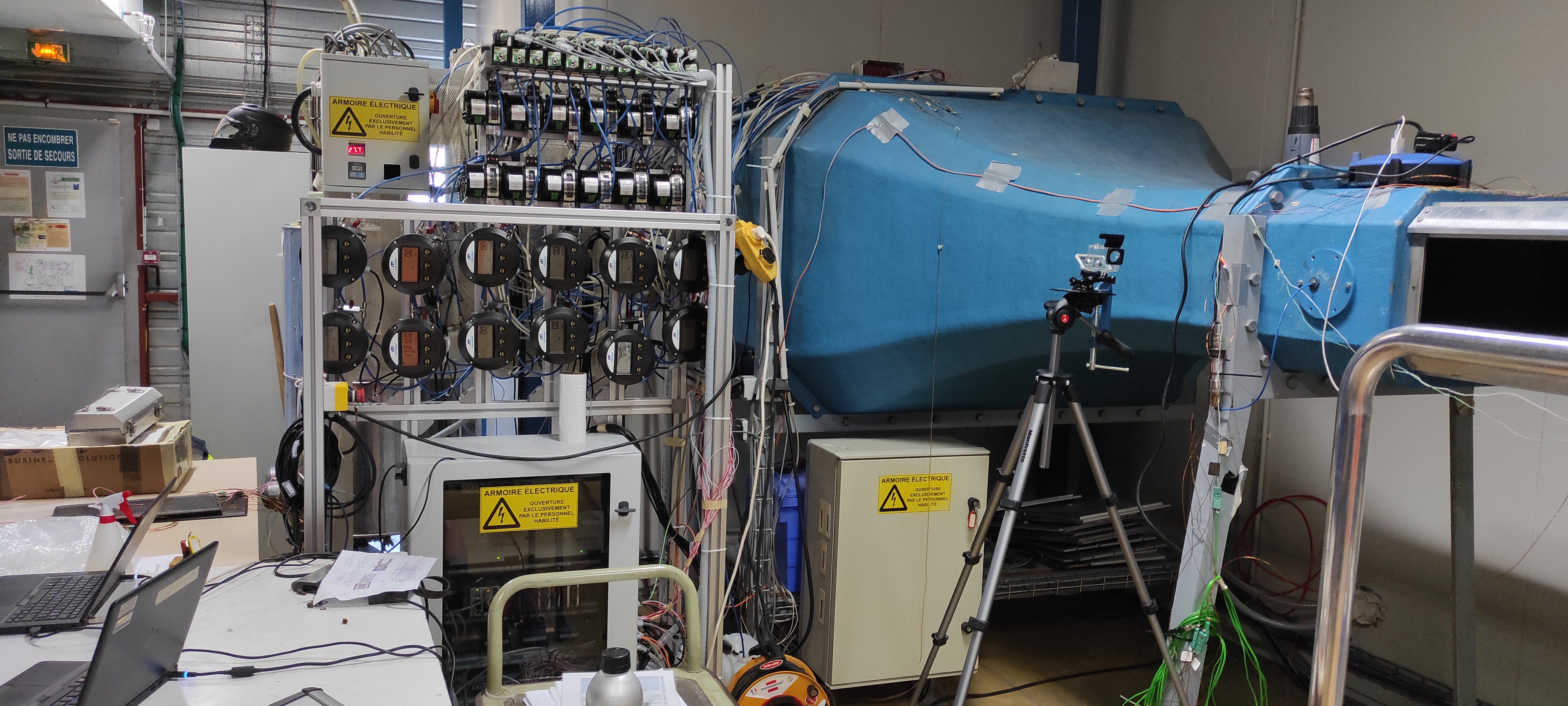}
		\caption{Flowmeters and control valves.}
		\label{fig:flowmeter_control_valve}
	\end{subfigure}
	\begin{subfigure}[b]{1\textwidth}
		\centering
		\includegraphics[width=\textwidth]{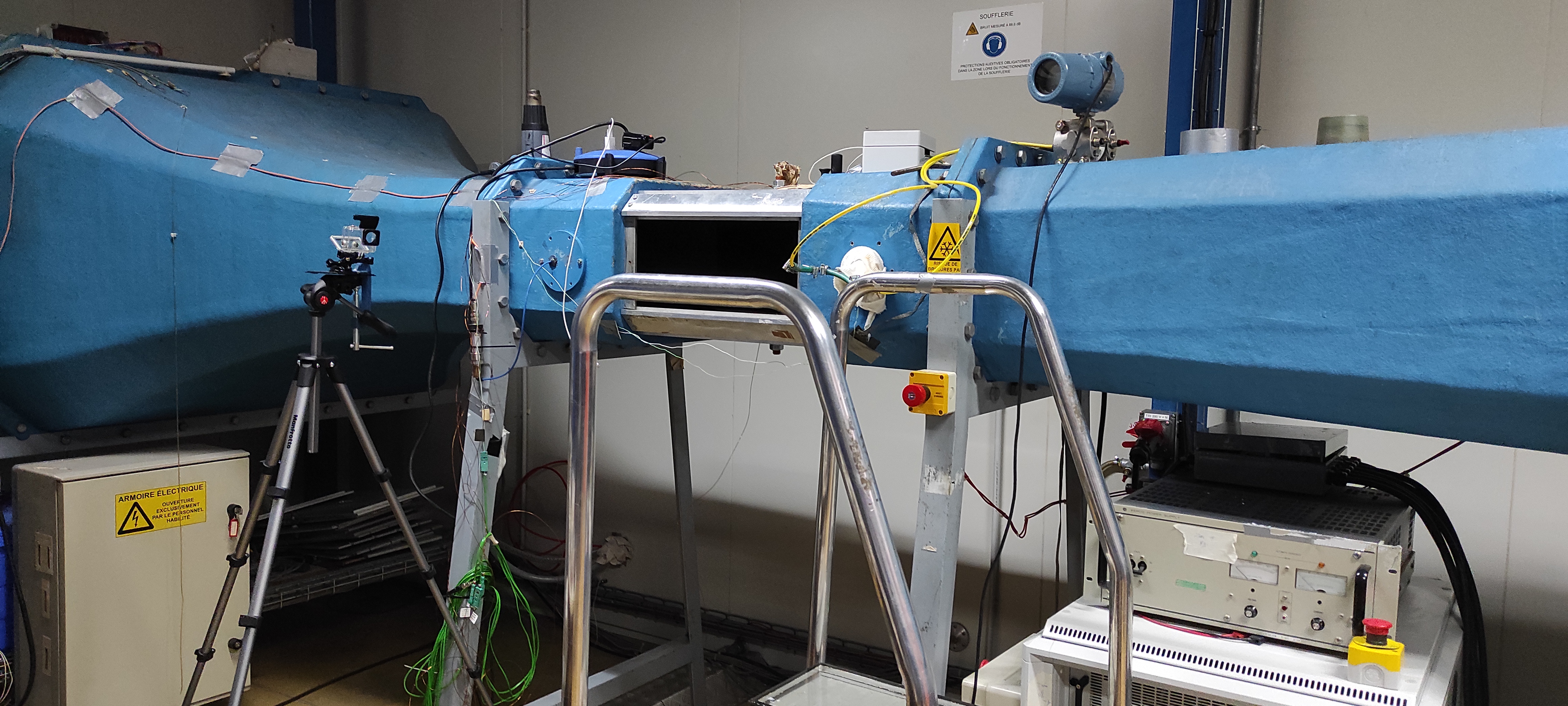}
		\caption{Test section.}
		\label{fig:Test section}
	\end{subfigure}
  \end{minipage}
  \hfill
  \begin{minipage}{0.48\columnwidth}
	\begin{subfigure}[b]{1\textwidth}
		\centering
		\includegraphics[width=0.8\textwidth]{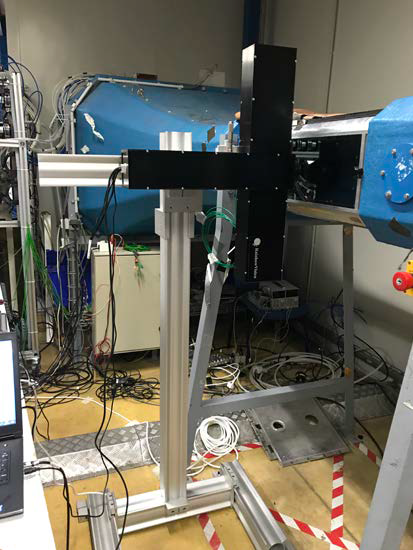}
		\caption{JRT measuring instrument.}
		\label{fig:jrt_instrument}
	\end{subfigure}
  \end{minipage}
  \caption{Wind tunnel facilities.}
  \label{fig:wind tunnel facilities}
\end{figure}

As mentioned above, each tank is connected to 12 conduits in which the flows are controlled by solenoid proportional valves.
The 12 \emph{air valves} are of type ASCO SCG202A001V and the 12 \emph{water valves} are of type ASCO SCG202A051V.
These valves are controlled by proportional integral (PI) controllers.
Furthermore, each bus is equipped with flowmeters that can be used to measure air and water flows, as shown in Figure \ref{fig:flowmeter_control_valve}.

The droplets are accelerated and injected into the test section by \emph{nozzles} of type SUJ16 pneumatic atomizers.
Each nozzle has a heating patch that is used to maintain the temperature above $\SI{0}{\degreeCelsius}$ to prevent the water droplets from freezing before injection in the test section.

Figure \ref{fig:Test section} shows the front part of the \emph{test section}.
In this area, the velocity and temperature of the wind are controlled throughout the experiment by external control systems.

As detailed previously, the entire plant is used to test defrost systems.
Therefore, it is necessary to control as accurately as possible the LWC and MVD of water droplets injected to perfectly characterize the tested devices.
For the present works, this means that we require a precise model giving these LWC and MVD as a function of the injection system's parameters.
To do so, in previous experiments, a JRT measuring instrument was placed in front of the test section (see Figure \ref{fig:jrt_instrument}) to accurately measure LWC and MVD.
The next section describes the data set obtained in these experiments.

		\subsection{Experimentation and description of the data collected}
			\label{subsection:data_collected}

To obtain LWC and MVD data for modelling, thirty experiments were carried out.
Data were collected from sensors installed throughout the plant as described in the previous section.
The data collected during these experiments are
\begin{itemize}
	\item the temperatures in the test section $T_{\text{TS}}$, the water tank $T_{\text{w}}$, the air tank $T_{\text{a}}$, and the nozzles $T_{\text{n}}$, all in Celsius degrees;
	\item the wind velocity in the test section $v_{\text{TS}}$ in $\si{\meter\per\second}$;
	\item the LWC $\Lambda$ in the test section in $\si{\gram\per\meter\cubed}$;
	\item the MVD $\mathrm{M}$ in the test section in $\si{\micro\meter}$;
	\item the volumetric air flow per bus $Q_{\text{a}}$ in $\si{\liter\per\minute}$ and volumetric water flow per bus $Q_{\text{w}}$ in $\si{\liter\per\hour}$.
\end{itemize}
The liquid droplets produced by the atomisers have a size of $\num{15}$ to $\SI{50}{\micro\meter}$ and are injected at a speed of $\num{25}$ to $\SI{50}{\meter\per\second}$.
In the test section, these droplets are supercooled at temperatures from $\num{0}$ to $\SI{-15}{\degreeCelsius}$.
In such experiments, we obtain LWC values ranging from $\num{0.1}$ to $\SI{2.5}{\gram\per\meter\cubed}$, while MVD values range between $\num{16.9}$ and $\SI{48.8}{\micro\meter}$.
Table \ref{table:statistical_indices} shows the main statistical indices of these experiments, calculated over thirty values for each variable, and Table \ref{table:correlation_variables} shows the correlation between the variables involved.

\begin{table*}[!tb]
	\centering
	\caption{Experimental statistical indices.}
	\label{table:statistical_indices}
	\begin{tabular}{c|ccccccccc}
				& $T_{\text{TS}}$ & $T_{\text{w}}$ & $T_{\text{a}}$ & $T_{\text{n}}$ & $v_{\text{TS}}$ & $\Lambda$ & $\mathrm{M}$ & $Q_{\text{a}}$ & $Q_{\text{w}}$\\
		\midrule
		mean               & \num{-6.5}  & \num{70.4} & \num{70.7} & \num{38.7} & \num{34.2} & \num{1.4}  & \num{32.7} & \num{23.7} & \num{9.0}  \\
		standard deviation & \num{2.9}   & \num{1.0}  & \num{1.8}  & \num{10.1} & \num{12.3} & \num{0.9}  & \num{9.7}  & \num{17.6} & \num{2.8}  \\
		minimum            & \num{-15.0} & \num{67.0} & \num{68.0} & \num{24.0} & \num{25.0} & \num{0.3}  & \num{16.9} & \num{10.0} & \num{4.5}  \\
		\SI{25}{\percent}  & \num{-7.6}  & \num{70.0} & \num{70.0} & \num{31.6} & \num{25.0} & \num{0.6}  & \num{25.4} & \num{10.0} & \num{8.1}  \\
		\SI{50}{\percent}  & \num{-6.0}  & \num{70.1} & \num{70.0} & \num{37.8} & \num{25.0} & \num{1.0}  & \num{30.9} & \num{15.0} & \num{8.1}  \\
		\SI{75}{\percent}  & \num{-5.0}  & \num{71.0} & \num{70.0} & \num{42.0} & \num{50.0} & \num{2.5}  & \num{42.8} & \num{30.0} & \num{11.2} \\
		maximum            & \num{-1.2}  & \num{73.0} & \num{74.0} & \num{70.0} & \num{50.0} & \num{2.5}  & \num{48.8} & \num{55.0} & \num{13.5} \\
	\end{tabular}
\end{table*}

\begin{table*}[!tb]
	\centering
	\caption{Correlation between variables.}
	\label{table:correlation_variables}
	\begin{tabular}{c|ccccccccc}
				& $T_{\text{TS}}$ & $T_{\text{w}}$ & $T_{\text{a}}$ & $T_{\text{n}}$ & $v_{\text{TS}}$ & $\Lambda$ & $\mathrm{M}$ & $Q_{\text{a}}$ & $Q_{\text{w}}$ \\
		\midrule
		$T_{\text{TS}}$ & \num{1.00}  & \num{-0.18} & \num{-0.05} & \num{0.36}  & \num{-0.16} & \num{0.39}  & \num{-0.11} & \num{0.19}  & \num{0.25}  \\
		$T_{\text{w}}$  & \num{-0.18} & \num{1.00}  & \num{0.24}  & \num{0.23}  & \num{-0.04} & \num{0.20}  & \num{0.08}  & \num{-0.20} & \num{0.09}  \\
		$T_{\text{a}}$  & \num{-0.05} & \num{0.24}  & \num{1.00}  & \num{-0.25} & \num{0.23}  & \num{0.03}  & \num{0.24}  & \num{-0.33} & \num{0.03}  \\
		$T_{\text{n}}$  & \num{0.36}  & \num{0.23}  & \num{-0.25} & \num{1.00}  & \num{-0.01} & \num{0.48}  & \num{0.08}  & \num{-0.07} & \num{0.47}  \\
		$v_{\text{TS}}$ & \num{-0.16} & \num{-0.04} & \num{0.23}  & \num{-0.01} & \num{1.00}  & \num{-0.06} & \num{-0.16} & \num{0.18}  & \num{0.35}  \\
		$\Lambda$       & \num{0.39}  & \num{0.20}  & \num{0.03}  & \num{0.48}  & \num{-0.06} & \num{1.00}  & \num{-0.10} & \num{-0.01} & \num{0.75}  \\
		$\mathrm{M}$    & \num{-0.11} & \num{0.08}  & \num{0.24}  & \num{0.08}  & \num{-0.16} & \num{-0.10} & \num{1.00}  & \num{-0.83} & \num{-0.03} \\
		$Q_{\text{a}}$  & \num{0.19}  & \num{-0.20} & \num{-0.33} & \num{-0.07} & \num{0.18}  & \num{-0.01} & \num{-0.83} & \num{1.00}  & \num{0.02}  \\
		$Q_{\text{w}}$  & \num{0.25}  & \num{0.09}  & \num{0.03}  & \num{0.47}  & \num{0.35}  & \num{0.75}  & \num{-0.03} & \num{0.02}  & \num{1.00}  \\
	\end{tabular}
\end{table*}

The data collected in these former experiments is then used in the following section to get a numerical model usable for control and simulation. 

	\section{Modelling methodology}
		\label{section:modelling_methodology}

The development of mathematical models in icing wind tunnels is highly complex due to the large number of variables and complex processes involved.
In this section, a hybrid nonlinear model is developed based on lumped parameters approach and machine learning techniques.
With this strategy, we provide a mathematical representation for the water and air tanks, the flow control valves, the nozzles, and the test section, aiming to link the LWC and MVD with these components' parameters.

		\subsection{Water tank}
			\label{subsection:water_tank_modelling}

\begin{figure}[!tb]
	\begin{subfigure}[b]{0.48\columnwidth}
		\centering
		\includegraphics[width=\textwidth]{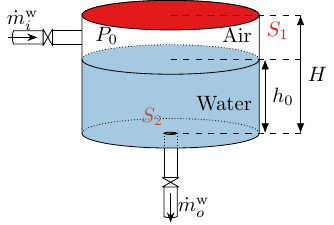}
		\caption{Water tank at initial state.}
		\label{fig:water_tank_initial_state}
	\end{subfigure}
	\hfill
	\begin{subfigure}[b]{0.48\columnwidth}
		\centering
		\includegraphics[width=\textwidth]{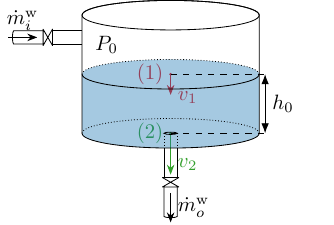}
		\caption{Water tank at instant $t$.}
		\label{fig:water_tank_instant_t}
	\end{subfigure}
	\caption{Water tank.}
	\label{fig:water_tank}
\end{figure}

As presented in Section \ref{subsection:equipment_description}, one of the first elements of the plant we study in this paper is a water tank.
To study the dynamic behaviour of this water tank, illustrated in Figure \ref{fig:water_tank}, we split the modelling into two parts.
On the one hand, \emph{mass balance} is used to model the level of water inside the tank.
On the other hand, \emph{energy balance} is used to model the influence of heat sources on the temperature of the water.

\paragraph{Water level}

The water tank has a cross-sectional area $S_1$, in $\si{\meter\squared}$, and a height $H$, in $\si{\meter}$.
As mentioned in the previous section, the tank is pressurised, at all times, with a constant pressure $P_0 = \SI{7}{\bar}$.
At the bottom of the tank is an opening having a cross-sectional area $S_2$, in $\si{\meter\squared}$.
When water is injected into the test section, this opening allows the tank to empty into the opened water buses.
At this moment, the surface of the water in the tank flows down at a velocity $v_1$, while it flows at a velocity $v_2$ through the orifice, both velocities being in $\si{\meter\per\second}$.
Finally, when emptying, the height of the liquid will evolve from an initial value $h_0$, in $\si{\meter}$, to adopt a value $h(t)$, in $\si{\meter}$.

The mass balance law provides the relation
\begin{equation} \label{eq:mass_balance_water_tank}
	\frac{dm}{dt} = \dot{m}^{\text{w}}_i - \dot{m}^{\text{w}}_o
\end{equation}
where $m$ is the water mass inside the tank, and $\dot{m}^{\text{w}}_i$ and $\dot{m}^{\text{w}}_o$ are the input and output fluids mass flow rate inside the tank.

\begin{rmk}
	In the following, the input mass flow rate $\dot{m}^{\text{w}}_i$ is a control variable that is chosen by the user depending on the operational needs.
	In addition, $\dot{m}^{\text{w}}_o$ is the water mass flow rate injected into the test section.
\end{rmk}

In the tank, the overall mass can be divided into a mass of water $m_{\text{w}}$ and a mass of air $m_{\text{wa}}$, so that
\begin{equation}
	\frac{dm_{\text{w}}}{dt} + \frac{dm_{\text{wa}}}{dt} = \dot{m}^{\text{w}}_i - \dot{m}^{\text{w}}_o\text{.}
\end{equation}
However, given the tank's volume, we have $m_{\text{w}} \gg m_{\text{wa}}$.
It follows that the mass of the air can be considered constant, so that
\begin{equation}
	\frac{dm_{\text{w}}}{dt} = \dot{m}^{\text{w}}_i - \dot{m}^{\text{w}}_o\text{.}
\end{equation}
As described before, the tank is cylindrical in shape and $m_{\text{w}} = \rho_{\text{w}}S_1h$, with $\rho_{\text{w}}$ the water density in $\si{\kilo\gram\per\meter\cubed}$, so that
\begin{equation}
	\frac{d(\rho_{\text{w}}S_1h)}{dt} = \dot{m}^{\text{w}}_i - \dot{m}^{\text{w}}_o\text{.}
\end{equation}
Therefore
\begin{equation}
	\frac{dh}{dt} = \frac{\dot{m}^{\text{w}}_i - \dot{m}^{\text{w}}_o}{\rho_{\text{w}}S_1}
\end{equation}
where $\rho_{\text{w}}$, in $\si{\kilo\gram\per\meter\cubed}$, is the water density.

\paragraph{Water temperature}

In terms of temperature, the energy change in volume control is due to several elements.
The first source of change is the power provided by the heating system $Q_{\text{w}}$, in $\si{\watt}$.
The second source are the losses caused by leaks to the outside of system $-\kappa_{\text{w}}T_{\text{w}}$ where $\kappa_{\text{w}}$, in $\si{\watt\per\degreeCelsius}$, is a constant characterising the leaks, and $T_{\text{w}}$, in $\si{\degreeCelsius}$, is the temperature in the water tank.
The third source is the input energy provided by the incoming mass $m_iC_eT^{\text{w}}_i$, where $C_e$, in $\si{\joule\per\kilo\gram\per\kelvin}$, is the water's mass specific heat, and $T^{\text{w}}_i$, in $\si{\kelvin}$, is the temperature of the incoming water.
Finally, the last source is the output energy provided by the outgoing mass $m_oC_eT^{\text{K}}_{\text{w}}$, where $T^{\text{K}}_{\text{w}} = T_{\text{w}} + 273.15$ is the temperature in the water tank expressed in Kelvin.
Then, considering the energy $E$, in $\si{\joule}$, of liquid water at a certain temperature, we have
\begin{equation}
	E = m_{\text{w}}C_eT^{\text{K}}_{\text{w}}
\end{equation}

The energy rate of change is then
\begin{equation} \label{eq:energy_change}
	\frac{dE}{dt} = \frac{dm_{\text{w}}}{dt}C_eT^{\text{K}}_{\text{w}} + m_{\text{w}}C_e\frac{dT_{\text{w}}}{dt}
\end{equation}
and, using \eqref{eq:mass_balance_water_tank} into \eqref{eq:energy_change},
\begin{equation}
	\frac{dE}{dt} = (\dot{m}^{\text{w}}_i - \dot{m}^{\text{w}}_o)C_eT^{\text{K}}_{\text{w}} + m_{\text{w}}C_e\frac{dT_{\text{w}}}{dt}\text{.}
\end{equation}
Given what we have described above, we have
\begin{equation}
	\frac{dE}{dt} = C_e(\dot{m}^{\text{w}}_iT^{\text{w}}_i - \dot{m}^{\text{w}}_oT^{\text{K}}_{\text{w}}) + Q_{\text{w}} - \kappa_{\text{w}}T_{\text{w}}
\end{equation}
so that
\begin{multline}
	(\dot{m}^{\text{w}}_i - \dot{m}^{\text{w}}_o)C_eT^{\text{K}}_{\text{w}} + mC_e\frac{dT_{\text{w}}}{dt}\\
		= \dot{m}^{\text{w}}_iC_eT^{\text{w}}_i - \dot{m}^{\text{w}}_oC_eT^{\text{K}}_{\text{w}} + Q_{\text{w}} - \kappa_{\text{w}}T_{\text{w}}
\end{multline}
or
\begin{equation}
	\frac{dT_{\text{w}}}{dt} = \frac{\dot{m}^{\text{w}}_iC_e(T^{\text{w}}_i - T^{\text{K}}_{\text{w}}) + Q_{\text{w}} - \kappa_{\text{w}}T_{\text{w}}}{m_{\text{w}}C_e}\text{.}
\end{equation}

		\subsection{Air tank}
			\label{subsection:air tank modelling}

The second element of the plant described in Section \ref{subsection:equipment_description} is the air tank.
In this Section, this air tank is described as a blower.
It follows that its geometry is not relevant for modelling purposes, and the important elements are the density of the air blown and its temperature.

As mentioned in \cite{ilic2016model}, following the law of conservation of matter \cite{anderson1990modern,oosthuizen2013introduction}, the air tank mass variation is equal to the mass flow inlet $\dot{m}^{\text{a}}_i$ minus the mass flow injected into the test section $\dot{m}^{\text{a}}_o$.
In addition, the mass of air in the tank $m_{\text{a}}$ can be written $m_{\text{a}} = \rho_{\text{a}}V_{\text{AT}}$ where $\rho_{\text{a}}$, in $\si{\kilo\gram\per\meter\cubed}$ is the air density, and $V_{\text{AT}}$, in $\si{\meter\cubed}$, is the volume of the air tank.
It follows that the variation of air density in the air tank is
\begin{equation}
	\frac{d\rho_{\text{a}}}{dt} = \frac{\dot{m}^{\text{a}}_i - \dot{m}^{\text{a}}_o}{V_{\text{AT}}}\text{.}
\end{equation}

With regard to temperature, the same methodology that has been used for the water tank has been followed.
Therefore, the evolution of the air temperature in the air tank is
\begin{equation}
	\frac{dT_{\text{a}}}{dt} = \frac{\dot{m}^{\text{a}}_i C_p(T^{\text{a}}_i - (T_{\text{a}} + 273.15)) + Q_{\text{a}} - \kappa_{\text{a}}T_{\text{a}}}{m_{\text{a}}C_p}
\end{equation}
where $T_{\text{a}}$, in $\si{\degreeCelsius}$, is the temperature of air in the air tank, $T^{\text{a}}_i$, in $\si{\kelvin}$, the temperature of the air injected inside the tank, $Q_{\text{a}}$, in $\si{\watt}$, the power provided by the heating system, $C_p$ is the air's mass specific heat, in $\si{\joule\per\kilo\gram\per\kelvin}$, and $\kappa_{\text{a}}$, in $\si{\watt\per\degreeCelsius}$, a constant characterizing the leaks.

		\subsection{Flow control valves}
			\label{subsection:valves modelling}

The elements placed right after the water and air tank are flow control valves allowing to control the flow of the mix of air and water injected into the test section of the plant.
As mentioned in Section \ref{subsection:equipment_description}, these water and air control valves are proportional solenoid valves.
Table \ref{table:valve_specifications} shows the main characteristics of these valves.

\begin{table}[!tb]
	\centering
	\caption{Proportional solenoid valve specifications.}
	\label{table:valve_specifications}
	\begin{tabular}{lc}
		\toprule
			\textbf{Orifice size $(\si{\milli\meter})$} & $\num{1.2}$\\
			\textbf{Flow factor} $\boldsymbol{K_v}$ ($\si{\meter\cubed\per\hour}$) & $\num{0.05}$\\
			\textbf{Operation pressure} ($\si{\bar}$) & $[0, 16]$\\
		\bottomrule
	\end{tabular}
\end{table}

\paragraph{Water valve}

According to the technical report \cite{technicalValves} of the water valve provided by the manufacturer, the volumetric flow of water through a valve $Q_{\text{wv}}$ is a function of the valve's opening area $A$, in $\si{\meter\squared}$, the discharge coefficient $C_d$, the pressure difference $\Delta P_{\text{wv}}$, in $\si{\pascal}$, in the valve, and the density of water $\rho_{\text{w}}$, in $\si{\kilo\gram\per\meter\cubed}$, and is given by
\begin{equation}
	Q_{\text{wv}} = AC_d\sqrt{2\frac{\Delta P_{\text{wv}}}{\rho_{\text{w}}}}\text{.}
\end{equation}
The discharge coefficient $C_d$ is obtained as
\begin{equation}
	C_d = \frac{4K_v}{\pi D^2} \sqrt{\frac{\rho_{\text{w}}}{2}}
\end{equation}
where $K_v$ is the flow factor, in $\si{\meter\cubed\per\hour}$, given by the manufacturer, and $D$, in $\si{\meter}$, is the diameter of the valve's orifice.
The difference of pressure $\Delta P_{\text{wv}}$ is obtained as
\begin{equation}
	\Delta P_{\text{wv}} = \frac{1}{2}\xi \rho_{\text{w}}v_{\text{wv}}^2
\end{equation}
where $\xi$ is the pressure drop coefficient and $v_{\text{wv}}$, in $\si{\meter\per\second}$, is the velocity at which water flows through the valve.
The pressure drop coefficient $\xi$ is given by
\begin{equation}
	\xi = \frac{\pi D^4}{8000K_v^2}
\end{equation}

\paragraph{Air valve}

Through the air valve, the volumetric flow $Q_{\text{av}}$ is given by \cite{ilic2017cascade,von2001unsteady}
\begin{equation}
	Q_{\text{av}} = \frac{A}{\rho_{\text{a}}}\sqrt{\frac{2\gamma}{R(\gamma-1)}}\frac{P_{\text{a}}}{\sqrt{T_{\text{a}}}}\left (\frac{P_{\text{av}}}{P_{\text{a}}}\right)^{\frac{1}{\gamma}}\sqrt{1-\left(\frac{P_{\text{av}}}{P_{\text{a}}} \right)^{\frac{\gamma-1}{\gamma}}}
\end{equation}
where $\gamma = 1.4$ is the ratio of specific heats, $R = \SI{2870}{\joule\per\kilo\gram\per\kelvin}$ is the air-gas constant, $P_{\text{a}}$, in $\si{\pascal}$, is the air pressure in the air tank pressure, and $P_{\text{av}}$, in $\si{\pascal}$, is the air pressure outside the control flow valve.

		\subsection{Nozzles}
			\label{subsection:nozzles}

Downstream the flow control valves we just described are nozzles at which level the air and water coming from the tanks are mixed and injected into the test section.
Nozzles are elements that are used to increase wind velocity based on their geometry.
As shown in Figure \ref{fig:testbed}, twelve nozzles are installed at the entrance of the test section.
They are arranged in three columns of four nozzles each.
This Section aims to study the air velocity that each nozzle provides to the test section, as well as the temperature in the nozzles.
A temperature higher than zero must be maintained so that there is no freezing in this part and the experiments can occur safely.
As can be seen in Table \ref{table:statistical_indices}, temperatures between $24$ and $\SI{70}{\degreeCelsius}$ are maintained in this area.

\paragraph{Velocity at the nozzles' output}

\begin{figure}[!tb]
	\centering
	\includegraphics[width=0.55\columnwidth]{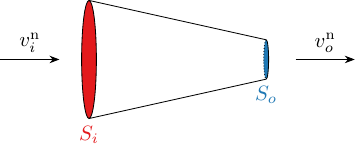}
	\caption{Nozzle's geometry.}
	\label{fig:nozzle_geometry}
\end{figure}

With regard to \emph{velocity}, a very simplified model using \emph{Bernoulli's continuity equation} is used to determine the velocity provided by the nozzles towards the \emph{test section}.
Bernoulli's equation can be viewed as a conservation of energy law for a flowing fluid.
Figure \ref{fig:nozzle_geometry} presents a schematic view the nozzles' geometry.
The input surface $S_i$ is taken $1.2$ times bigger than the outlet surface $S_o$.
Hence, the velocity at the outlet of the nozzles is given by
\begin{equation}
	S_iv^{\text{n}}_i = S_ov^{\text{n}}_o
\end{equation}
where $S_i$ and $S_o$, in $\si{\meter\squared}$, are respectively the inlet and outlet surfaces, and $v^{\text{n}}_i$ and $v^{\text{n}}_o$, in $\si{\meter\per\second}$, are the velocity of the air coming from the air tank and of the mix of air and water injected in the test section.
It follows that
\begin{equation}
	v^{\text{n}}_{o} = 1.2 v^{\text{n}}_{i}.
\end{equation}

\begin{rmk}
	As usual for this type of equipment \cite{knop2021comparison}, the velocity of the water injected inside the nozzle is not taken into account, as it is negligible.
	It is the pressure provoked by the air velocity that leads to the spraying of a mix of air and water at the output of the nozzle.
\end{rmk}

\paragraph{Data-driven model of the mix's temperature in the nozzles}

Temperatures in the nozzles $T_{\text{n}}$ are not measured by a sensor.
However, it can be estimated by obtaining a mathematical model that depends on other known variables.
To do so, we performed a statistical data analysis to identify variables that have a greater impact on the temperature of the mixture of air and water in the nozzles.
We then extract a polynomial model of $T_{\text{n}}$ depending on the identified variable.

\begin{figure}[!tb]
	\begin{subfigure}[b]{.9\columnwidth}
		\centering
		\includegraphics{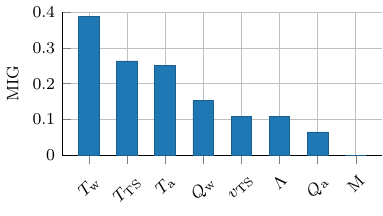}
		\caption{$T_{\text{n}}$ mutual information gain.}
		\label{fig:tn_mig}
	\end{subfigure}
	\begin{subfigure}[b]{.9\columnwidth}
		\centering
		\includegraphics{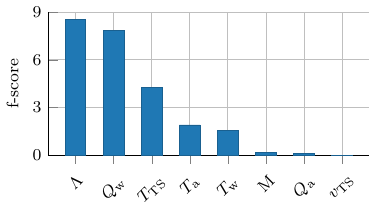}
		\caption{$T_{\text{n}}$ f-score.}
		\label{fig:tn_f_score}
	\end{subfigure}
	\caption{MIG and f-score for $T_{\text{n}}$.}
	\label{fig:tn_indices}
\end{figure}

To determine which of the system's variables can be used to express $T_{\text{n}}$, we use \emph{mutual information gain (MIG)} \cite{vinh2012novel} and \emph{f-score} \cite{goutte2005probabilistic}.
Figure \ref{fig:tn_mig} presents the mutual information gain of the system's variables with $T_{\text{n}}$.
It shows that the temperature of the nozzles depends on other temperature values, where the most important variables are $T_{\text{w}}$, $T_{\text{TS}}$, and $T_{\text{a}}$, followed by $Q_{\text{w}}$, $v_{\text{TS}}$, $\Lambda$, and $Q_{\text{a}}$ that have a minor influence on the temperature of the mixture in the nozzle $T_{\text{n}}$.
In addition, according to the f-score, presented in Figure \ref{fig:tn_f_score}, $\Lambda$ and $Q_{\text{w}}$ have a greater influence than the variables $T_{\text{w}}$, $T_{\text{TS}}$, and $T_{\text{a}}$.
Therefore, to ensure that the nozzle temperature only depends on other temperature values, we chose not to use $\Lambda$ and $Q_{\text{w}}$ to estimate $T_{\text{n}}$.
As a result, the variables $T_{\text{w}}$, $T_{\text{TS}}$, and $T_{\text{a}}$ have been used to obtain different models based on the data for the evolution of $T_{\text{n}}$.

From this analysis, a regression tree, a neural network \cite{rodriguez2015machine}, and a polynomial model \cite{tso2007predicting} for $T_{\text{n}}$ are computed and compared with each other.
The mix's temperature in the nozzles $T_{\text{n}}$ is given as a function of only $T_{\text{w}}$, $T_{\text{TS}}$, and $T_{\text{a}}$.
As described in Section \ref{subsection:data_collected}, a low number of experiments ($30$ in total) were performed, leading to a low number of data.
It follows that all acquired data has to be used to estimate the models.
In order to avoid overfitting, this is done using a cross-validation of $5$.
The models are obtained by minimising the mean square error (MSE), \cite{chai2014root}.
Finally, the models are validated on the $30$ experiments themselves.

\begin{figure}[!tb]
	\centering
	\includegraphics[width=0.8\columnwidth]{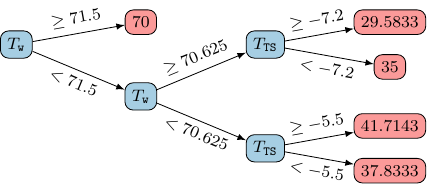}
	\caption{Regression tree for $T_{\text{n}}$.}
	\label{fig:tnozzle_regression_tree}
\end{figure}

A simple form is imposed to determine the polynomial model giving $T_{\text{n}}$ as a function of $T_{\text{w}}$, $T_{\text{TS}}$, and $T_{\text{a}}$.
Therefore, minimizing the MSE, we obtain
\begin{equation}
	T_{\text{n}} = -79.5664 + 1.4220T_{\text{TS}} + 3.6303T_{\text{w}} - 1.8113T_{\text{a}}
\end{equation}
where all the temperatures are expressed in degree Celsius.
The regression tree giving the value of $T_{\text{n}}$ is presented in Figure \ref{fig:tnozzle_regression_tree}.
It has to be read as tests on the variables $T_{\text{w}}$, $T_{\text{TS}}$, and $T_{\text{a}}$, expressed in degree Celsius.
Depending on the position of these variables with given thresholds, indicated on the tree's branches, we obtain a value, in degree Celsius, for $T_{\text{n}}$.

For the neural model of $T_{\text{n}}$, we optimize the properties of the neural network among the following possibilities:
\begin{itemize}
	\item the number of hidden layers is either $1$, $2$, or $3$;
	\item each hidden layer can contain between $2$ and $10$ neurons;
	\item the activation function is either $sigmoid$, $tanh$, or $relu$.
\end{itemize}
As for the other models, a 5-fold cross-validation is used and the characteristics of the neural network are obtained through minimisation of the MSE.
The optimal characteristics of the neural model of $T_{\text{n}}$ are presented in Table \ref{table:tn_nn_structure}.

\begin{table}[!tb]
	\centering
	\caption{$T_{\text{n}}$ neural network structure.}
	\label{table:tn_nn_structure}
	\begin{tabular}{lcccccc}
		\toprule
		\textbf{Hidden layers} & 2\\ 
		\textbf{Neurons per layer} & 10 \& 4\\
		\textbf{Activation function} & sigmoid\\
		\bottomrule
	\end{tabular}
\end{table}

\begin{figure}[!tb]
	\begin{subfigure}[b]{.9\columnwidth}
		\centering
		\includegraphics{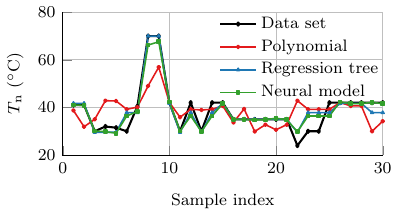}
		\caption{$T_{\text{n}}$ validation.}
		\label{fig:tnozzle_validation}
	\end{subfigure}
	\begin{subfigure}[b]{.9\columnwidth}
		\centering
		\includegraphics{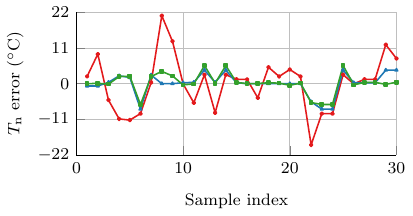}
		\caption{$T_{\text{n}}$ error.}
		\label{fig:tnozzle_error}
	\end{subfigure}
	\caption{$T_{\text{n}}$ validation and error.}
	\label{fig:tnozzle_validation_error}
\end{figure}

The proposed models are first validated on the data from previous experiments described in Section \ref{subsection:data_collected}.
To do so, Figure \ref{fig:tnozzle_validation} compares the values of $T_{\text{n}}$ obtained with the polynomial model (in red), the regression tree (in blue), and the neural model (in green) with the values of $T_{\text{n}}$ from the 30 experiments in the data set.
We see that the regression tree and the neural model behave better than the polynomial model.
Figure \ref{fig:tnozzle_error} confirms this behaviour as it shows the error between each model and the data from the experiments, the colours being the same as in Figure \ref{fig:tnozzle_validation}.
To further validate the models, we present the overall mean square error and mean absolute error (MAE) \cite{chai2014root} for the different $T_{\text{n}}$ models in Table \ref{table:tnozzle_models_performance}.
It confirms that the regression tree and the neural model perform better than the polynomial model, with the neural model performing slightly better than the tree.
However, due to the low quantity of data available in the data set, the regression tree shown in Figure \ref{fig:tnozzle_regression_tree} has a low number of possible states for $T_{\text{n}}$.
Therefore, we decide not to use it in the following as it might not work well in a more general simulation process.

\begin{table}[!tb]
	\centering
	\caption{$T_{\text{n}}$ models performance.}
	\label{table:tnozzle_models_performance}
	\begin{tabular}{lcccccc}
		\toprule
			\textbf{Model}           & \textbf{MSE} & \textbf{MAE}\\
		\midrule
			\textbf{Polynomial}      & 66.70 & 6.20 \\
			\textbf{Regression tree} & 10.68 & 2.03 \\
			\textbf{Neural network}  & 9.57  & 1.97 \\
		\bottomrule
	\end{tabular}
\end{table}

		\subsection{Test section}
			\label{subsection:test_section}

As already exposed, the goal of the plant is to inject a mix of air and water into the test section.
As explained in Section \ref{subsection:equipment_description}, the variables of importance are the liquid water content $\Lambda$ and the median volumetric diameter $\mathrm{M}$.
Two other variables, the temperature $T_{\text{TS}}$ and the wind velocity $v_{\text{TS}}$ in the test section are also important.
However, they are regulated by external systems, so we consider them as input variables and focus on LWC and MVD modelling.

\paragraph{Liquid water content}

According to \cite{knop2021comparison}, LWC $\Lambda$ represents the mixing of the available mass of water $m^{\text{w}}_{\text{TS}}$ within a defined air volume $V^{\text{a}}_{\text{TS}}$ in the test section, given by
\begin{equation}
	\Lambda = \frac{m^{\text{w}}_{\text{TS}}}{V^{\text{a}}_{\text{TS}}}\text{.}
\end{equation}
The volume of air flowing through the test section can be calculated by $V^{\text{a}}_{\text{TS}} = v_{\text{TS}}A_{\text{TS}}$ where $A_{\text{TS}}$ is the cross-sectional area of the test section.
The available mass of water is $m^{\text{w}}_{\text{TS}} = \rho_{\text{w}}Q_{\text{w}}$.
It follows that
\begin{equation} \label{eq:lwc_expression}
	\Lambda = \frac{\rho_{\text{w}}Q_{\text{w}}}{v_{\text{TS}}A_{\text{TS}}}\text{.}
\end{equation}

\begin{figure}[!tb]
	\begin{subfigure}[b]{.9\columnwidth}
		\centering
		\includegraphics{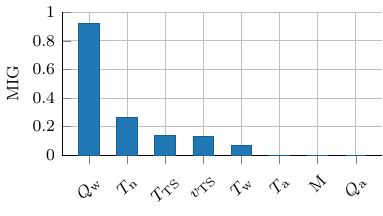}
		\caption{LWC mutual information gain.}
		\label{fig:lwc_mig}
	\end{subfigure}
	\begin{subfigure}[b]{.9\columnwidth}
		\centering
		\includegraphics{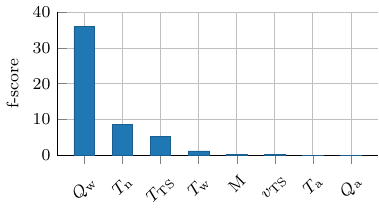}
		\caption{LWC f-score.}
		\label{fig:lwc_f_score}
	\end{subfigure}
	\caption{MIG and f-score for LWC.}
	\label{fig:lwc_indices}
\end{figure}

The MIG and f-score for the LWC, showing the influence of the plant's variables over the LWC, are presented in Figure \ref{fig:lwc_indices}.
They show that the flow of water $Q_{\text{w}}$ plays indeed an important role in the expression of $\Lambda$ as is implied by the relation \eqref{eq:lwc_expression}.
However, the influence of the air velocity in the test section $v_{\text{TS}}$ is less important according to both the MIG and f-score.
Nevertheless, the expression \eqref{eq:lwc_expression} is grounded into other experiments from other testbeds \cite{knop2021comparison}.
Therefore, we decide to model the LWC with the expression given in equation \eqref{eq:lwc_expression}.

\paragraph{Mean volume diameter}

As highlighted in the Introduction, the estimation of the MVD is a much more complicated task.
In the literature, various measurement devices \cite{knop2021comparison} or machine learning techniques \cite{rydblom2020measurement} have been used, as mentioned previously.
Therefore, from the data described in Section \ref{subsection:data_collected}, we follow the same methodology to model the MVD $\mathrm{M}$ as we did for the temperature in the nozzle $T_{\text{n}}$ as exposed in Section \ref{subsection:nozzles}.

\begin{figure}[!tb]
	\begin{subfigure}[b]{.9\columnwidth}
		\centering
		\includegraphics{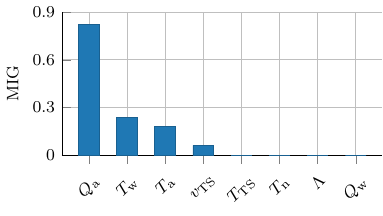}
		\caption{MVD mutual information gain.}
		\label{fig:mvd_mig}
	\end{subfigure}
	\begin{subfigure}[b]{.9\columnwidth}
		\centering
		\includegraphics{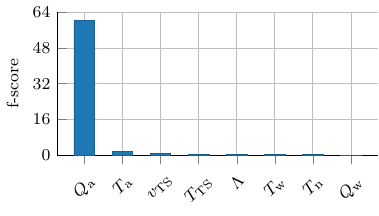}
		\caption{MVD f-score.}
		\label{fig:mvd_f_score}
	\end{subfigure}
	\caption{MIG and f-score for MVD.}
	\label{fig:mvd_indices}
\end{figure}

We start by obtaining the mutual information gain and f-score to know which variables have the greatest influence on the MVD's value.
The MIG and f-score for MVD are shown in Figure \ref{fig:mvd_indices}.
According to the MIG values in Figure \ref{fig:mvd_mig}, the variables $Q_{\text{a}}$, $T_{\text{w}}$, $T_{\text{a}}$, and $v_{\text{TS}}$ are the variables that influence the MVD value.
In addition, the variable $Q_{\text{a}}$ has a more important influence compared to the others.
This observation is in accordance with the f-score presented in Figure \ref{fig:mvd_f_score}.

Figures \ref{fig:lwc_indices} and \ref{fig:mvd_indices} show valuable information, as the LWC $\Lambda$ does not appear to have a great influence on the value of MVD $\mathrm{M}$ and vice versa.
However, because of the definition of these two variables, if the liquid water content in the droplets is zero, we cannot get a median volumetric diameter.
In addition, we already saw that LWC values depend on the water flow $Q_{\text{w}}$, and MVD values depend on the air flow $Q_{\text{a}}$.

It follows, by symmetry, and the influence that it has according to Figure \ref{fig:mvd_indices}, that we decide to express the MVD $\mathrm{M}$ as a function of $Q_{\text{a}}$.
From the previous discussion, we also decide to include the LWC $\Lambda$ in the MVD's expression.
Finally, although they do not have a great influence on the MVD according to the values in Figure \ref{fig:mvd_indices}, we include the variables describing the conditions in the test section, i.e., the temperature $T_{\text{TS}}$ and the wind velocity $v_{\text{TS}}$, as the MVD is measured in the test section.

As for the temperature of the mix of air and water in the nozzles $T_{\text{n}}$, we want to obtain a polynomial, a regression tree, and a neural model for the MVD.
The process for obtaining them is the same as described in Section \ref{subsection:nozzles}.

\begin{figure}[!tb]
	\centering
	\includegraphics[width=.8\columnwidth]{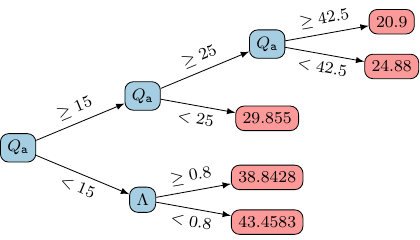}
	\caption{Regression tree for MVD.}
	\label{fig:mvd_regression_tree}
\end{figure}

To obtain the MVD polynomial, more complex forms than for $T_{\text{n}}$ are imposed.
First, as $Q_{\text{a}}$ has the highest influence on the MVD $\mathrm{M}$ according to Figure \ref{fig:mvd_indices}, we propose a polynomial model where $Q_{\text{a}}$ is a factor of all terms composed of all possible combinations of 1 to 3 variables among $\Lambda$, $T_{\text{TS}}$, and $v_{\text{TS}}$ to the power $1$.
Minimizing the MSE with this form, we obtain
\begin{multline} \label{eq:mvd_a}
	\mathrm{M} = 44.3881 - 0.6299Q_{\text{a}} + 0.0178Q_{\text{a}}v_{\text{TS}} + 0.0257Q_{\text{a}}T_{\text{TS}}\\
		- 0.1530Q_{\text{a}}\Lambda + 0.0012Q_{\text{a}}v_{\text{TS}}T_{\text{TS}} - 0.0035Q_{\text{a}}v_{\text{TS}}\Lambda\\
		- 0.0546Q_{\text{a}}T_{\text{TS}}\Lambda + 0.0003Q_{\text{a}}v_{\text{TS}}T_{\text{TS}}\Lambda
\end{multline}
The second form we consider involves all possible combinations of products of 1 to 4 variables among $Q_{\text{a}}$, $\Lambda$, $T_{\text{TS}}$, and $v_{\text{TS}}$ to the power $1$.
Therefore, minimizing the MSE, we obtain
\begin{multline} \label{eq:mvd_b}
	\mathrm{M} = - 7.6323T_{\text{TS}} + 0.8825v_{\text{TS}} + 8.8270\Lambda + 4.4380Q_{\text{a}}\\
		+ 0.1419v_{\text{TS}}T_{\text{TS}} + 0.4914T_{\text{TS}}\Lambda - 0.1066v_{\text{TS}}\Lambda\\
		+ 0.8393Q_{\text{a}}T_{\text{TS}} - 0.0812Q_{\text{a}}v_{\text{TS}} - 0.9817Q_{\text{a}}\Lambda\\
		+ 0.0189v_{\text{TS}}T_{\text{TS}}\Lambda - 0.0141Q_{\text{a}}T_{\text{TS}}v_{\text{TS}} - 0.111Q_{\text{a}}T_{\text{TS}}\Lambda\\
		- 0.0037Q_{\text{a}}v_{\text{TS}}\Lambda - 0.0023Q_{\text{a}}T_{\text{TS}}v_{\text{TS}}\Lambda
\end{multline}
where the temperature $T_{\text{TS}}$ is expressed in degree Celsius.
The regression tree giving the value of $\mathrm{M}$ is presented in Figure \ref{fig:mvd_regression_tree}.
It is interpreted in the same way that the regression tree for $T_{\text{n}}$ is interpreted as described in Section \ref{subsection:nozzles}.

The neural model of $\mathrm{M}$ is obtained under the same conditions as the neural model of $T_{\text{n}}$.
Its optimal characteristics are presented in Table \ref{table:mvd_nn_structure}.

\begin{table}[!tb]
	\centering
	\caption{MVD neural network structure.}
	\label{table:mvd_nn_structure}
	\begin{tabular}{lcccccc}
	\toprule
		\textbf{Hidden layers}       & 2 \\
		\textbf{Neurons per layer}   & 2 \& 7\\
		\textbf{Activation function} & ReLu\\
	\bottomrule
	\end{tabular}
\end{table}

\begin{figure}[!tb]
	\begin{subfigure}[b]{0.9\columnwidth}
		\centering
		\includegraphics{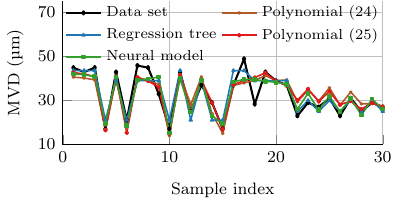}
		\caption{MVD validation.}
		\label{fig:mvd_validation}
	\end{subfigure}
	\begin{subfigure}[b]{0.9\columnwidth}
		\centering
		\includegraphics{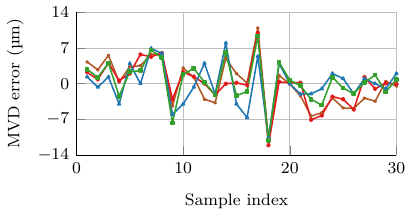}
		\caption{MVD error.}
		\label{fig:mvd_error}
	\end{subfigure}
	\caption{MVD validation and error.}
\label{fig:mvd_validation_error}
\end{figure}

As for the temperature $T_{\text{n}}$, the proposed models are first validated on the data from previous experiments described in Section \ref{subsection:data_collected}.
To do so, Figure \ref{fig:mvd_validation} compares the MVD values $\mathrm{M}$ obtained with the polynomial model from \eqref{eq:mvd_a} (in brown), the polynomial model from \eqref{eq:mvd_b} (in red), the regression tree (in blue), and the neural model (in green) with the values of $\mathrm{M}$ from the 30 experiments in the data set.
From the curve, it is difficult to properly separate the different models as they have quite a similar behaviour.
Figure \ref{fig:mvd_error} confirms that as it shows the error between each model and the data from the experiments, the colours being the same as in Figure \ref{fig:mvd_validation}.
To further validate the models, we present the overall mean square error and mean absolute error for the different $\mathrm{M}$ models in Table \ref{table:mvd_models_performance}.
It confirms that the models behave really closely.
However, with the values of Table \ref{table:mvd_models_performance}, we decide not to use the polynomial model given by equation \eqref{eq:mvd_a} as it clearly has poorer performance compared to the other models.
As for the temperature in the nozzle $T_{\text{n}}$, we also decide not to use the regression tree, we also decide not to consider the regression tree model given in Figure \ref{fig:mvd_regression_tree} as, due to the low number of data in the data set, gives a low number of possible states.

\begin{table}[!tb]
	\centering
	\caption{MVD models performance.}
	\label{table:mvd_models_performance}
	\begin{tabular}{lcccccc}
		\toprule
		\textbf{Models}          & \textbf{MSE} & \textbf{MAE} \\
		\midrule
		\textbf{Polynomial \eqref{eq:mvd_a}} & $\num{20.70}$ & $\num{3.71}$ \\ 
		\textbf{Polynomial \eqref{eq:mvd_b}} & $\num{17.47}$ & $\num{2.84}$ \\
		\textbf{Regression tree} 			 & $\num{16.50}$ & $\num{3.10}$ \\
		\textbf{Neural model}    			 & $\num{16.79}$ & $\num{3.11}$ \\
	\bottomrule
	\end{tabular}
\end{table}

	\section{Results}
		\label{section:results}

\begin{figure*}[!tb]
	\centering
	\includegraphics[width=1\textwidth]{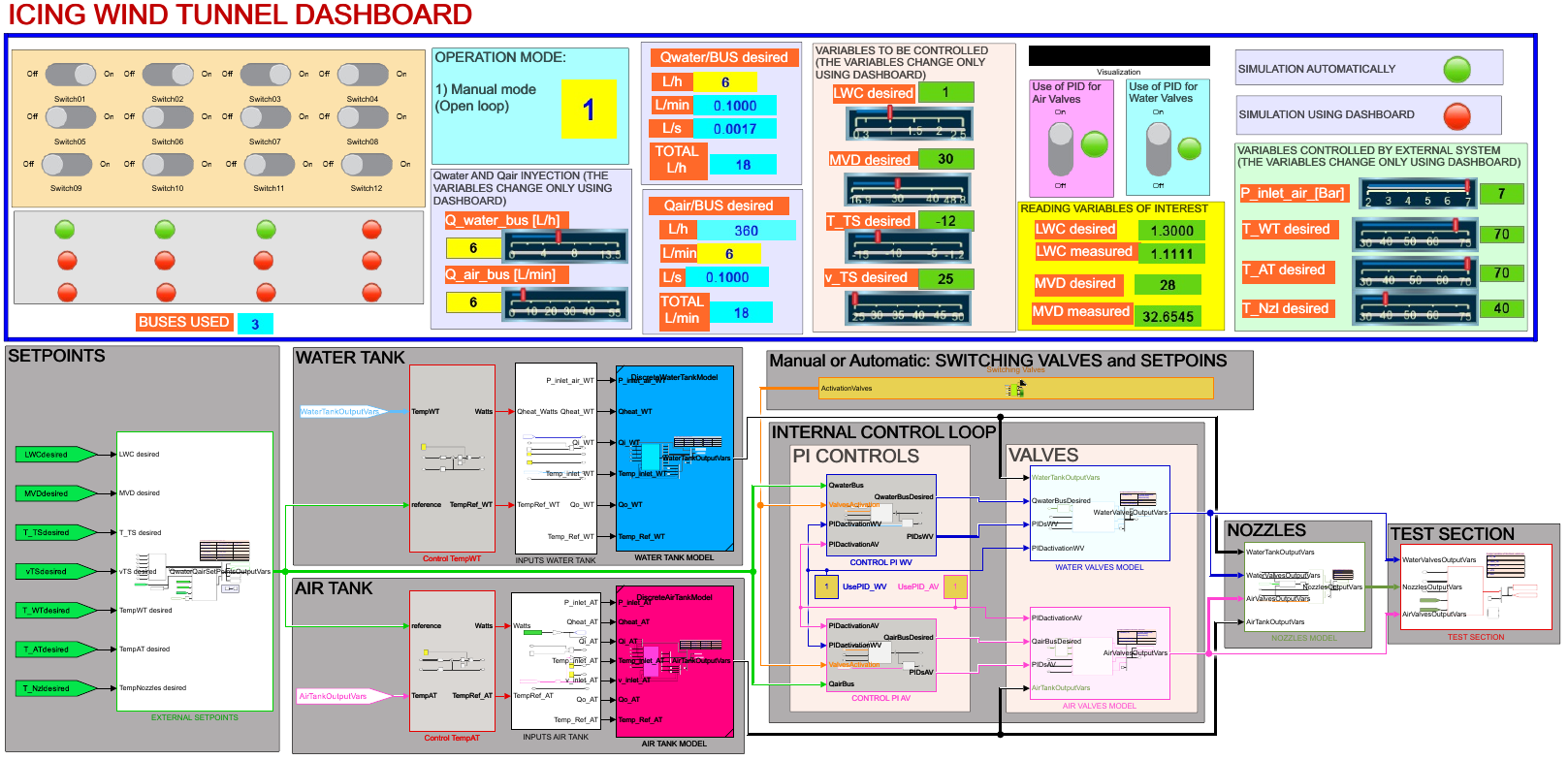}
	\caption{Modelling Simulink diagram.}
	\label{fig:simulink_diagram}
\end{figure*}

In this section, we present results obtained in simulation and discuss the analysis and behaviour of the system.
As mentioned previously, it is of great importance to have atmospheric conditions that encourage the study of variables such as LWC and MVD that are related to ice formation.
The models developed in Section \ref{section:modelling_methodology} are incorporated in interconnected modules in a Simulink model shown in Figure \ref{fig:simulink_diagram}, which formed a hybrid model of the plant. 
A dashboard with different functionalities has been designed to simulate scenarios in which various modes of plant operation can be used.
The results shown in this section are from a simulation of $\SI{20}{\minute}$ with a sampling time of $\SI{1}{\second}$.
The conditions considered in the simulation are described below.

\begin{itemize}
	\item The variables involved in the simulation process have been executed within the ranges of real conditions as shown in Table \ref{table:statistical_indices}.
	\item Different valves have been activated and deactivated throughout the experiment.
 	\item In the simulation, changes in the temperature of the test section, as well as in the velocity of the test section, have been considered. 
	\item The temperature ($T_{\text{TS}}$) and the velocity ($v_{\text{TS}}$) in the test section are considered variables controlled by an external system.
	\item The flow through each valve is controlled by a PI controller (in the internal control loop) and the PI acts on the opening and closing area of the valves.
\end{itemize}

\begin{figure*}[!tb]
	\begin{subfigure}[h]{0.5\textwidth}
		\centering
		\includegraphics{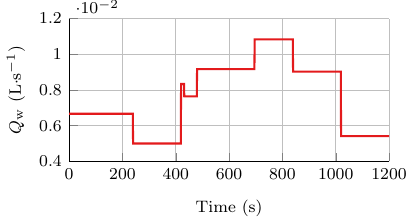}
		\caption{Water flow $Q_{\text{w}}$ in open loop.}
		\label{fig:LWC_action_results_open_loop}	
	\end{subfigure}
	\begin{subfigure}[h]{0.5\textwidth}
		\centering
		\includegraphics{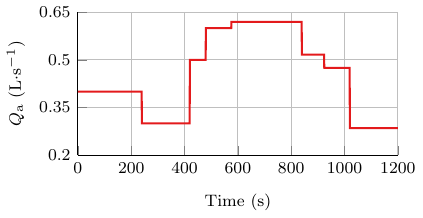}
		\caption{Air flow $Q_{\text{a}}$ in open loop.}
		\label{fig:MVD_action_results_open_loop}	
	\end{subfigure}
	\begin{subfigure}[h]{0.5\textwidth}
		\centering
		\includegraphics{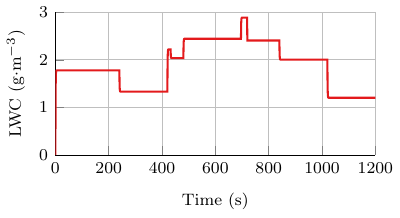}
		\caption{LWC time response in open loop.}
		\label{fig:LWC_results_open_loop}
	\end{subfigure}
	\begin{subfigure}[h]{0.5\textwidth}
		\centering
		\includegraphics{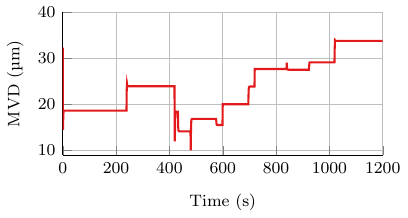}
		\caption{MVD time response in open loop.}
		\label{fig:MVD_results_open_loop}
	\end{subfigure}
	\begin{subfigure}[h]{0.5\textwidth}
		\centering
		\includegraphics{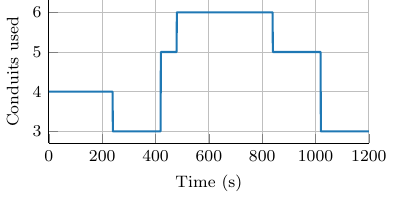}
		\caption{Conduits used.}
		\label{fig:conduits_used_open_loop}
	\end{subfigure}
	\begin{subfigure}[h]{0.5\textwidth}
		\centering
		\includegraphics{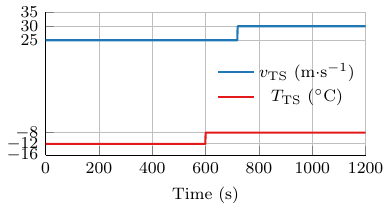}
		\caption{Velocity and temperature in test section.}
		\label{fig:v_TS_T_TS_open_loop}
	\end{subfigure}
	\caption{Time response for LWC and MVD in open loop.}
	\label{fig:LWC_actions_open_loop}
\end{figure*}

To observe the time response of LWC, a water flow $Q_{\text{w}}$ of $\SI{6}{\liter\per\hour}$ from initial time to $\SI{442}{\second}$, $\SI{5.5}{\liter\per\hour}$ from second $\SI{443}{\second}$ to $\SI{696}{\second}$, and $\SI{6.5}{\liter\per\hour}$ from $\SI{697}{\second}$ to $\SI{1200}{\second}$, respectively, per conduit (see Figure \ref{fig:LWC_action_results_open_loop}) has been applied.
LWC is directly proportional to $Q_{\text{w}}$ and inversely proportional to $v_{\text{TS}}$ according to the equation \eqref{eq:lwc_expression}. 
Therefore, the changes in LWC shown in Figure \ref{fig:LWC_results_open_loop} are due to the amount of water flow, the number of conduits used (Figure \ref{fig:conduits_used_open_loop}) and the changes in velocity in the test section (Figure \ref{fig:v_TS_T_TS_open_loop}).

MVD varies according to the equation \eqref{eq:mvd_b} where $Q_{\text{a}}$ has the strongest influence.
As is done to LWC, an air flow per conduit of $\SI{6}{\liter\per\minute}$ from the initial time to $\SI{576}{\second}$, $\SI{6.2}{\liter\per\minute}$ from $\SI{577}{\second}$ to $\SI{924}{\second}$, and $\SI{5.7}{\liter\per\minute}$ from $\SI{925}{\second}$ to $\SI{1200}{\second}$ has been applied during the simulation, as shown in Figure \ref{fig:MVD_action_results_open_loop}. 
The MVD dynamics is shown in Figure \ref{fig:MVD_results_open_loop}, it can be observed that for the values of air flow, and the variation of variables such as the number of conduits used, the velocity and temperature in the test section, as well as the LWC, the MVD values between $10$ and $\SI{35}{\micro\meter}$ can be obtained. 

\begin{figure*}[!tb]
	\begin{subfigure}[h]{0.5\textwidth}
		\centering
		\includegraphics{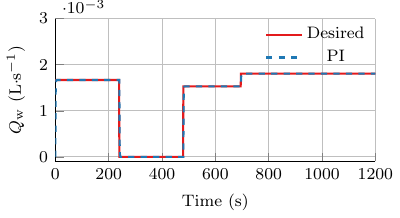}
		\caption{Behaviour of water valve 1.}
		\label{fig:Behaviour_WV_01_open_loop}
	\end{subfigure}
	\begin{subfigure}[h]{0.5\textwidth}
		\centering
		\includegraphics{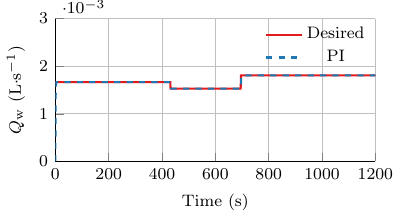}
		\caption{Behaviour water valve 2.}
		\label{fig:Behaviour_WV_02_open_loop}
	\end{subfigure}
	\begin{subfigure}[h]{0.5\textwidth}
		\centering
		\includegraphics{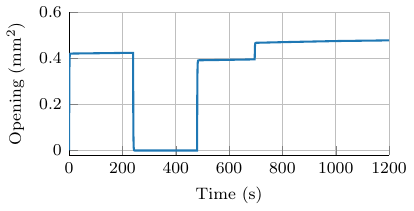}
		\caption{Opening water valve 1.}
		\label{fig:Opening_WV_01_open_loop}
	\end{subfigure}
	\begin{subfigure}[h]{0.5\textwidth}
		\centering
		\includegraphics{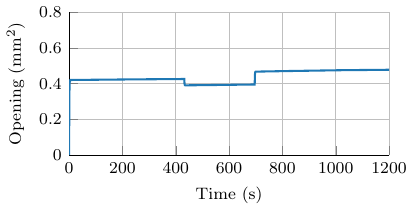}
		\caption{Opening water valve 2.}
		\label{fig:Opening_WV_02_open_loop}
	\end{subfigure}
	\begin{subfigure}[h]{0.5\textwidth}
		\centering
		\includegraphics{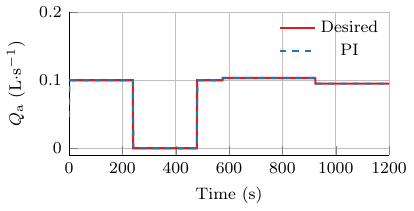}
		\caption{Behaviour air valve 1.}
		\label{fig:Behaviour_AV_01_open_loop}
	\end{subfigure}
	\begin{subfigure}[h]{0.5\textwidth}
		\centering
		\includegraphics{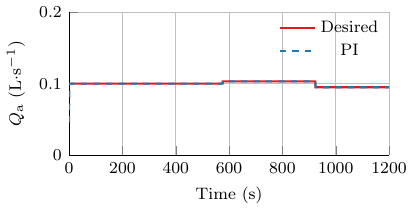}
		\caption{Behaviour air valve 2.}
		\label{fig:Behaviour_AV_02_open_loop}
	\end{subfigure}
	\begin{subfigure}[h]{0.5\textwidth}
		\centering
		\includegraphics{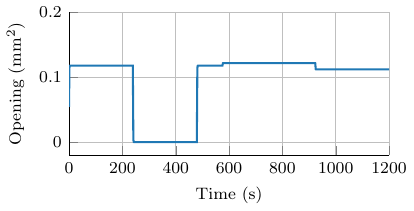}
		\caption{Opening air valve 1.}
		\label{fig:Opening_AV_01_open_loop}
	\end{subfigure}
	\begin{subfigure}[h]{0.5\textwidth}
		\centering
		\includegraphics{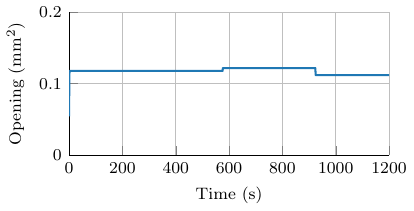}
		\caption{Opening air valve 2.}
		\label{fig:Opening_AV_02_open_loop}
	\end{subfigure}
	\caption{Water and air valves behaviour in open loop.}
	\label{fig:Water_and_valves_behaviour_open_loop}
\end{figure*}

Figures \ref{fig:Behaviour_WV_01_open_loop} and \ref{fig:Behaviour_WV_02_open_loop} show the flow of water through valves $1$ and $2$, it can be seen how the PI control maintains the desired flow through them. 
Figures \ref{fig:Opening_WV_01_open_loop} and \ref{fig:Opening_WV_02_open_loop} show the opening in $\si{\milli\meter}$ of the valves $1$ and $2$ through which the water flow passes. 
Furthermore, the behaviour of air through the valves $1$ and $2$ is shown in Figures \ref{fig:Behaviour_AV_01_open_loop} and \ref{fig:Behaviour_AV_02_open_loop}, respectively. 
Whereas the opening of these valves is shown in Figures \ref{fig:Opening_AV_01_open_loop} and \ref{fig:Opening_AV_02_open_loop}. 
The valve $1$ (water and air valves) are imposed an off and on behaviour at $\SI{240}{\second}$ and $\SI{480}{\second}$, respectively.

The dynamics of the whole system is highly complex, involving a large number of variables; however, the developed model allows us to study our system under different scenarios and allows us to achieve values close to experimental data. 

	\section{Conclusions}
		\label{section:conclusions}

In this paper, mathematical modelling based on first principles and machine learning techniques has been developed to form a hybrid model that allows us to perform an in-depth study of the atmospheric conditions in icing structures, using a subsonic wind tunnel as a facility.
A statistical analysis of experimental data has been carried out in order to have a global idea of the behaviour of the system, which has allowed us to develop machine learning models of some variables of interest in the plant under study. 
The implemented model has been validated with real data and delivers valuable results. 

The proposed methodology has allowed us to establish the dynamics of LWC and MVD in a simulation environment, as well as allowing us to make several variations in the parameters involved in the model, in order to study in depth the atmospheric conditions suitable for ice formation. 

In future work, we intend to use optimisation techniques in order to optimise parameters in the system and improve the performance. 
In addition, we intend to develop intelligent control strategies and advanced control strategies such as fuzzy control, reinforcement learning, and model predictive control that allow us to control the desired LWC and MVD values minimising the use of system resources. 

	\section*{Acknowledgements}

The authors would like to thank Vincent Sircoulomb, Associate Professor at ESIGELEC, for his help on the COPOGIRT project.

\bibliographystyle{IEEEtran}
\bibliography{model}

\end{document}